\documentclass[12pt,preprint]{aastex}

\usepackage{graphicx}
\usepackage{amsmath}

\newcommand{\eqb}{\begin{equation}}
\newcommand{\eqe}{\end{equation}}

\begin{document}

\title{Three-Dimensional Relativistic Magnetohydrodynamic Simulations of Current-Driven Instability. II. Relaxation of Pulsar Wind Nebula}

\author{
Yosuke Mizuno\altaffilmark{1,2}, Yuri Lyubarsky\altaffilmark{3}, Ken-Ichi
Nishikawa\altaffilmark{1,2}, and Philip E. Hardee\altaffilmark{4}}

\altaffiltext{1}{Center for Space
Plasma and Aeronomic Research, University of Alabama in Huntsville
, 320 Sparkman Drive, NSSTC, Huntsville, AL 35805, USA; mizuno@cspar.uah.edu}
\altaffiltext{2}{National Space Science and Technology Center,
VP62, Huntsville, AL 35805, USA}
\altaffiltext{3}{Physics Department, Ben-Gurion University, Beer-Sheva 84105, Israel}
\altaffiltext{4}{Department of
Physics and Astronomy, The University of Alabama, Tuscaloosa, AL
35487, USA}

\shorttitle{3D RMHD Simulations of CD Instability II}
\shortauthors{Mizuno et al.}

\begin{abstract}

We have investigated the relaxation of a hydrostatic hot plasma column containing toroidal magnetic field by the Current-Driven (CD) kink instability as a model of pulsar wind nebulae. In our simulations the CD kink instability is excited by a small initial velocity perturbation and develops turbulent structure inside the hot plasma column. We demonstrate that, as envisioned by Begelman, the hoop stress declines and the initial gas pressure excess near the axis decreases. The magnetization parameter $\sigma$, the ratio of the Poynting to the kinetic energy flux, declines from an initial value of $0.3$ to about $0.01$ when the CD kink instability saturates. Our simulations demonstrate that axisymmetric models strongly overestimate the elongation of the pulsar wind nebulae. Therefore, the previous requirement for an extremely low pulsar wind magnetization can be abandoned. The observed structure of the pulsar wind nebulae do not contradict the natural assumption that  the magnetic energy flux still remains a good fraction of the total energy flux after dissipation of alternating fields.
\end{abstract}

\section{Introduction}

The pulsar wind nebulae (PWNe) may be considered as a relativistically hot bubble continuously pumped by an electron-positron plasma and magnetic field emanating from the pulsar. Pulsars lose their rotational energy predominantly by generating a highly magnetized, ultrarelativistic wind. The wind presumably terminates at a strong reverse shock and the shocked plasma inflates a bubble within the external medium. The synchrotron and inverse Compton radiation from the shocked plasma is observed from the radio to the gamma-ray band (see the review by Gaensler \& Slane 2006).

Close to the pulsar the energy is carried mostly by electro-magnetic fields as Poynting flux; however, the common belief is that at the termination shock the wind must already be very weakly magnetized.
Simple spherical models of PWNe suggest that the magnetization parameter $\sigma$, the ratio of the Poynting to the kinetic energy flux, needs to be as small as 0.001-0.01 just upstream of the termination shock (Rees \& Gunn 1974; Kennel \& Coroniti 1984a,b; Emmering \& Chevalier 1987). The reason for the required low value of $\sigma$ at the termination shock is that conservation of the magnetic flux in the effectively incompressible subsonic flow downstream of the termination shock implies rapid increase in the magnetic field strength with radius and the field within the nebula could exceed the equipartition value if the magnetization at the termination shock is not extremely small. Extensive axisymmetric MHD simulations of the flow produced by the pulsar wind within a plerionic nebula (Komissarov \& Lyubarsky 2003, 2004; Del Zanna et al.\ 2004, 2006; Volpi et al.\ 2008; Camus et al.\ 2009) show that one can account for the morphology of PWNe, including the remarkable jet-torus structure, with $\sigma\approx 0.01$. If the magnetization were larger, the nebula would be elongated by the magnetic pinch effect beyond observational limits. Such a low value of $\sigma$ is puzzling because it is not easy to invent a realistic energy conversion mechanism to reduce $\sigma$ to the required level. This problem, often referred to as the ``$\sigma$ problem'' is widely discussed in the literature (see recent reviews by Arons 2007; Kirk et al.\ 2009).

One has to stress that all the available observation limits on $\sigma$ are obtained from the analysis of the plasma flow and radiation beyond the termination shock, where the upstream $\sigma$ is calculated from the ideal MHD jump conditions as if the Poyntning flux is transferred by large scale magnetic fields. However in the pulsar wind, most of the energy is transferred by waves, which an obliquely rotating magnetosphere excites near the light cylinder. These waves cannot propagate within the nebula because the wavelength (on the order of the light cylinder radius) is less than the particle Larmor radii. The above mentioned observational limits on $\sigma$ refer only to the mean magnetic field remaining after the oscillating part is erased.

In the equatorial belt of the wind, the sign of the magnetic field alternates with the pulsar period, forming stripes of opposite magnetic polarity (Michel 1971; Bogovalov 1999); such a structure is called a striped wind. In the striped wind, Poynting flux can be converted into particle energy flux when the oppositely directed magnetic fields annihilate. Observations of X-ray tori around pulsars (Gaensler \& Slane 2006) as well as theoretical modeling of the pulsar wind (Bogovalov 1999; Spitkovsky 2006) suggest that it is in the equatorial belt where most of the wind energy is transported. Therefore, in the equatorial belt magnetic dissipation of the striped wind is the main energy conversion mechanism in pulsars. It has been shown that due to relativistic time dilation, complete dissipation could occur only on a scale comparable to or larger than the radius of the termination shock (Lyubarsky \& Kirk 2001; Kirk \& Skjaeraasen 2003, Zenitani \& Hoshino 2007). However, the alternating fields still annihilate at the termination shock (Petri \& Lyubarsky 2007). At higher latitudes, where the magnetic field does not change sign, fast magnetosonic waves may transport a significant amount of energy. These waves can decay relatively easily (Lyubarsky 2003) but can release only a fraction of the Poynting flux into the plasma, because at these latitudes most of the energy is carried by the mean magnetic field.

The fraction of the energy transferred by the mean field can be found only from 3D numerical simulations of the pulsar magnetosphere. Even though this fraction is still not known, this fraction is less than 1/2 because the angular distribution of the Poynting flux in the pulsar wind is a maximum at the rotational equator, where the mean field is zero. This suggests that $\sigma$ becomes less than unity after the waves decay. Therefore, at a quantitative level the $\sigma$ problem is partially solved if Poynting flux is converted into plasma energy via dissipation of the oscillating part of the field. However, the residual $\sigma$ still cannot be as small as required by axisymmetric models. Therefore the question still remains as to how the mean field $\sigma$, which is only somewhat less than unity, could become extremely small.

Since no mechanism had been found for the extraction of energy from a large scale, axisymmetric magnetic field, Begelman (1998) suggested that the problem can be alleviated if a current-driven (CD) kink instability destroys the concentric field structure in the nebula. In the axisymmetric case, magnetic loops in the expanding flow are forced to expand and perform work against the magnetic tension. The CD kink instability allows the loops to come apart and one expects that in 3D, the mean field strength is not amplified much by expansion of the flow and the hoop stress would not necessarily pinch the flow as much as would otherwise be supposed. In this case $\sigma$ just upstream of the termination shock might not need to be so unreasonably small as was found in axisymmetric models.

This idea can be checked only by 3D simulations of plasma flow within the nebula. As a first step we simulate the 3D evolution of the simple cylindrical model of PWNe developed by Begelman \& Li (1992). This model describes a quasi-static cylindrical configuration with a purely toroidal magnetic field. The plasma within the cylinder is relativistically hot and the hoop stress is balanced by the thermal pressure. The cylinder is confined on the outside by a nonmagnetized gas. The linear analysis shows (Begelman 1998) that such a configuration is unstable with respect to the CD kink instability. Here we simulate the nonlinear evolution of this system in order to see whether stabilizing boundary conditions suppress development of the instability inside the plasma volume via 3D relativistic magnetohydrodynamic (RMHD) simulations of the CD kink instability in a relativistically hot plasma column containing a toroidal magnetic field. This paper is organized as follows: We describe the numerical method and setup used for our simulations in \S 2, present our results in \S 3, and discuss the astrophysical implications in \S 4.

\section{Numerical Setup}

To study the evolution of the CD kink instability in a hot plasma column containing a toroidal magnetic field, we use the 3D GRMHD code ``RAISHIN'' in three-dimensional Cartesian geometry. RAISHIN is based on a $3+1$ formalism of the general relativistic conservation laws of particle number and energy-momentum, Maxwell's equations, and Ohm's law with no electrical resistance (ideal MHD condition) in a curved spacetime (Mizuno et al.\ 2006)
\footnote{
Constained transport schemes are used to maintain divergence-free magnetic field in the RAISHIN code. This scheme requires the magnetic field to be defined at the cell interfaces. On the other hand, conservative, high-resolution shock capturing schemes (Godonov-type scheme) for conservation laws require the variables to be defined at the cell center. In order to combine variables defined at these different positions, the magnetic field calculated at the cell interfaces is interpolated to the cell center and as a result the scheme becomes non-conservative even though we solve the conservation laws (Komissarov 1999).}.
In the RAISHIN code, a high-resolution shock-capturing (HRSC) scheme is employed. The numerical fluxes are calculated using the HLL approximate Riemann solver, and flux-interpolated constrained transport (flux-CT) is used to maintain a divergence-free magnetic field. The RAISHIN code performs special relativistic calculations in Minkowski spacetime by choosing the appropriate metric. RAISHIN has proven to be accurate to second order and has passed a number of one-dimensional and multidimensional numerical tests including highly relativistic cases and highly magnetized cases in both special and general relativity (Mizuno et al.\ 2006; Mizuno et al.\ 2010).  Here a fifth-order weighted essentially non-oscillatory (WENO, Jiang \& Shu 1996) reconstruction scheme is built into the RAISHIN code and used in the simulations in order to handle turbulent structure more finely. WENO schemes provide highly-accurate solutions in regions of smooth flow and non-oscillatory transitions in the presence of discontinuous waves by combining different interpolation stencils of order $r$ into a weighted average of order $2r-1$.

We consider a hot plasma column containing a pure toroidal magnetic field with radius $R$ and the height $L_{z}$ being in hydrostatic equilibrium with a hot static medium. We begin with the cylindrical model for PWNe proposed by Begelman \& Li (1992) who calculated the axisymmetric structure of a bubble blown by a continuously injected relativistically hot, magnetized plasma in a surrounding medium. According to this model, the radial gas pressure and toroidal magnetic field profiles in the hot plasma column are given by
\begin{equation}
p={p_{0} \over \eta^2}, 
\end{equation}
\begin{equation}
{B^{2}_{\phi} \over 8 \pi}={ 9 p_{0} x^{2} \over 4 \eta^3}, 
\end{equation}
where $x=r/R$, and $\eta$ is found for any $x$ from the equation
\begin{equation}
\eta + {9 \over 8} x^{2} = \eta^{3/2} .
\end{equation}
In this solution, the magnetic hoop stress is balanced by the plasma pressure. Thus, the plasma pressure at the axis is larger than the total (plasma and magnetic) pressure at the edge of the bubble. It is this discrepancy that results in the elongation of the nebula. At $x > 1$, the hot plasma column is surrounded by a hot static unmagnetized medium with constant pressure equal to the total pressure at the bubble edge
\begin{equation}
p_{out}=\frac{2p_{0}}{\eta_{0}^{2}},
\end{equation}
where $\eta_{0}=9/4$ is the solution of Eq.\ (3) at $x=1$. We choose $\rho=1$ throughout the region. In the simulations,  we set $p_{0}=10^{5} \rho c^{2}$, so that the plasma is relativistically hot, i.e. $\rho c^{2} << p$, where $c$ is speed of light. The equation of state is that of an ideal gas with $p=(\Gamma-1)\rho e$, $e$ is the specific internal energy density and the adiabatic index is set to $\Gamma = 4/3$. The specific enthalpy is $h \equiv 1 + e/c^{2} + p/\rho c^{2}$.

The above configuration was shown to be neutrally stable with respect to axisymmetric perturbations but unstable to the helical modes (Begelman 1998). Therefore, we choose an initial small radial velocity perturbation with the form given by
\begin{equation}
v_{x,y}/c={\delta v \over c} e^{-r} \sum^{N}_{k=1} {a_{k;x,y} \over N} \sin \left( {2 \pi k z \over L_{z}} + \phi_{x} \right),
\end{equation}
where $N$ is the total number of modes, $\phi_{k}$ is a random phase selected from the range $0 < \phi_{k} < 2 \pi$, and $a_{k;x,y}$ is a random direction with $a_{k;x}^{2}+a_{k;y}^2=1$. We choose $\delta v=0.01c$ in the simulations. In order to evaluate the effect of different initial perturbations we consider two cases. In case A, we choose $N=2$, $\phi_{k}=0$, $(a_{1;x}, a_{1;y})=(1,0)$, and $(a_{2;x}, a_{2;y})=(0,1)$, and in case B, we choose $N=6$  and random phases and directions. The computational domain is a Cartesian ($x, y, z$) box of size $6R \times 6R \times  L_{z}$ ($L_{z} = R$) with grid resolution of $R=60 \Delta L$ and $L_{z}=60 \Delta L$. The grid resolution is the same in all directions. The simulation grid is periodic along the axial ($z$) direction. We use a larger computational domain in the $x$ and $y$ directions to avoid boundary effects. We impose reflecting boundary conditions in the $x$ and $y$ directions to ensure that the total energy is conserved.

\section{Results}

In Figure 1 we show the evolution of the magnetized column for case A when only a few modes are excited.  The overall evolution for the multiple modes case B is similar but structures are more complicated because more modes are excited.
\begin{figure}[hp!]
\epsscale{0.9}
\plotone{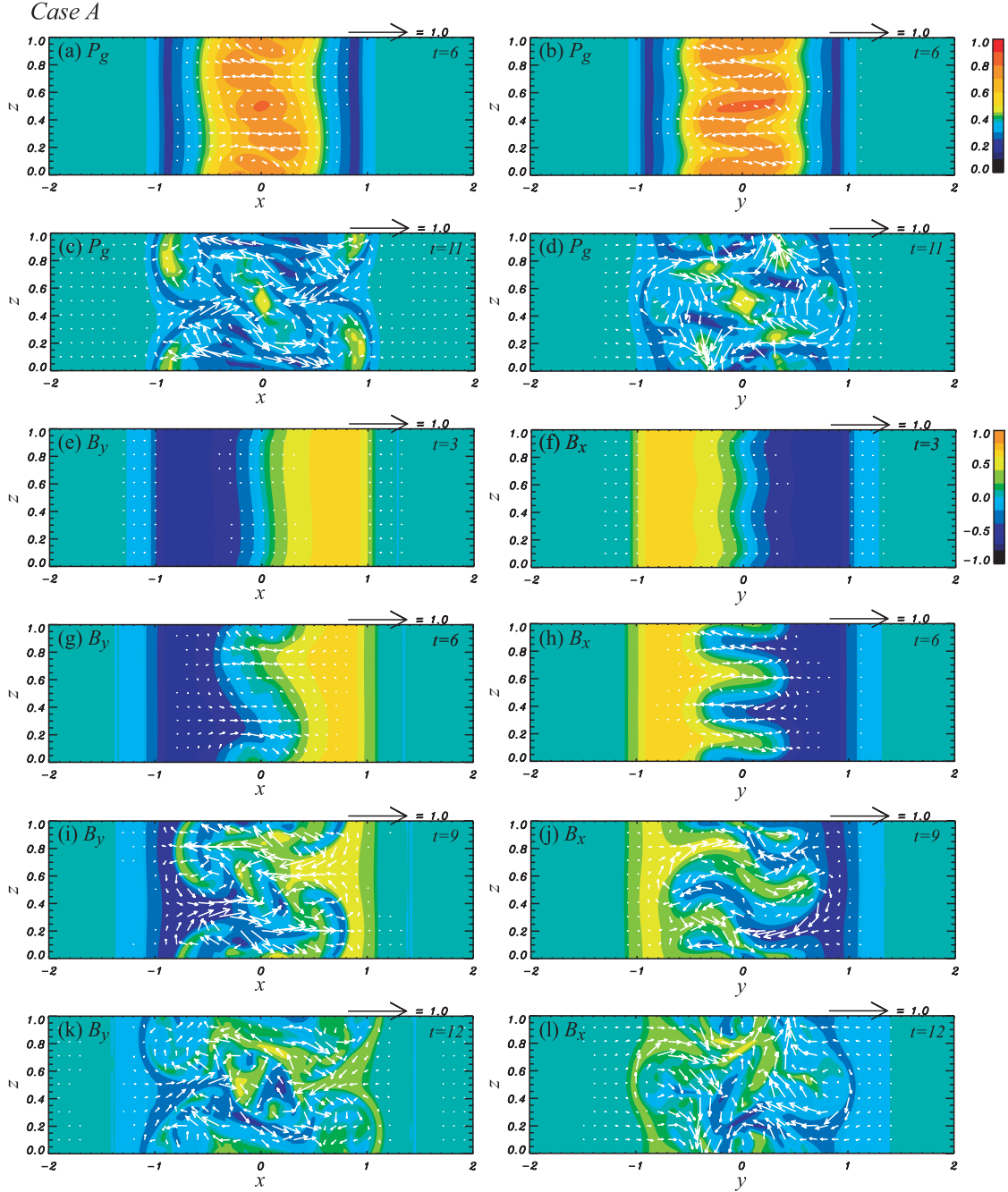}
\caption{Two dimensional images of (case A): the gas pressure in the $xz$ plane at $y=0$ for ({\it a}) $t=6 R/c$, ({\it c}) $11 R/c$, the gas pressure in the $yz$ plane at $x=0$ for ({\it b}) $t=6 R/c$, ({\it d}) $11 R/c$, the magnetic field $B_{y}$ in the $xz$ plane at $y=0$ for ({\it e}) $t=3 R/c$, ({\it g}) $6 R/c$, ({\it i}) $9 R/c$, ({\it k}) $12 R/c$, and $B_{x}$ in the $yz$ plane at $x=0$ for ({\it f}) $t=3 R/c$, ({\it h}) $6 R/c$, ({\it j}) $9 R/c$, ({\it l}) $12 R/c$. Arrows indicate the velocity in each plane. 
\label{f1}}
\end{figure}
The two-dimensional images of the gas pressure $p$ and the perpendicular components ($B_{y}$ and $B_{x}$) of the magnetic field in the $xz$ plane at $x=0$ and the $yz$ plane at $y=0$ are restricted to $R \le 2$ in order to focus on the strong interaction region. The initial small velocity perturbation excites the CD kink instability $n=1$ mode in the $x$-direction and $n=2$ mode in the $y$-direction (see Fig.\ 1 at $t=6R/c$). The radial velocity ($x$-component and $y$-component) increases with time in the linear growth phase. At about $t=6R/c$ the CD kink instability shifts to the non-linear phase (see also Fig.\ 2). In the non-linear phase, the two modes ($n=1$ and $n=2$) interact and lead to turbulence in the hot plasma column. The turbulent motion creates complicated magnetic field structures in the hot plasma column. Note that the gas pressure within the column, which was initially high in order to balance the magnetic hoop stress, decreases because the hoop stress weakens. Thus, as a result of the kink instability the magnetic loops come apart and release the magnetic stress. 
 
As an indicator of the growth of the CD kink instability, the time evolution of the volume-averaged plasma energy $E_{p}=\rho h \gamma^{2} -p$ and magnetic energy $E_{m}=B^{2}/2$ for cases A and B are shown in Figure 2. 
An initial slow evolution in the linear growth phase lasts up to $t=6R/c$ for case A ($t=7R/c$ for case B), and is followed by a more rapid evolution in a nonlinear growth phase. In the linear phase the magnetic energy gradually decreases with a following rapid decrease in the nonlinear phase. In the nonlinear phase, this rapid decrease of the magnetic energy ceases at about $t=11R/c$ for case A ($t=14R/c$ for case B). The same trend was seen in our previous study of the CD kink instability of a static cold plasma column (Mizuno et al.\ 2009). 
\begin{figure}[h!]
\epsscale{0.8}
\plotone{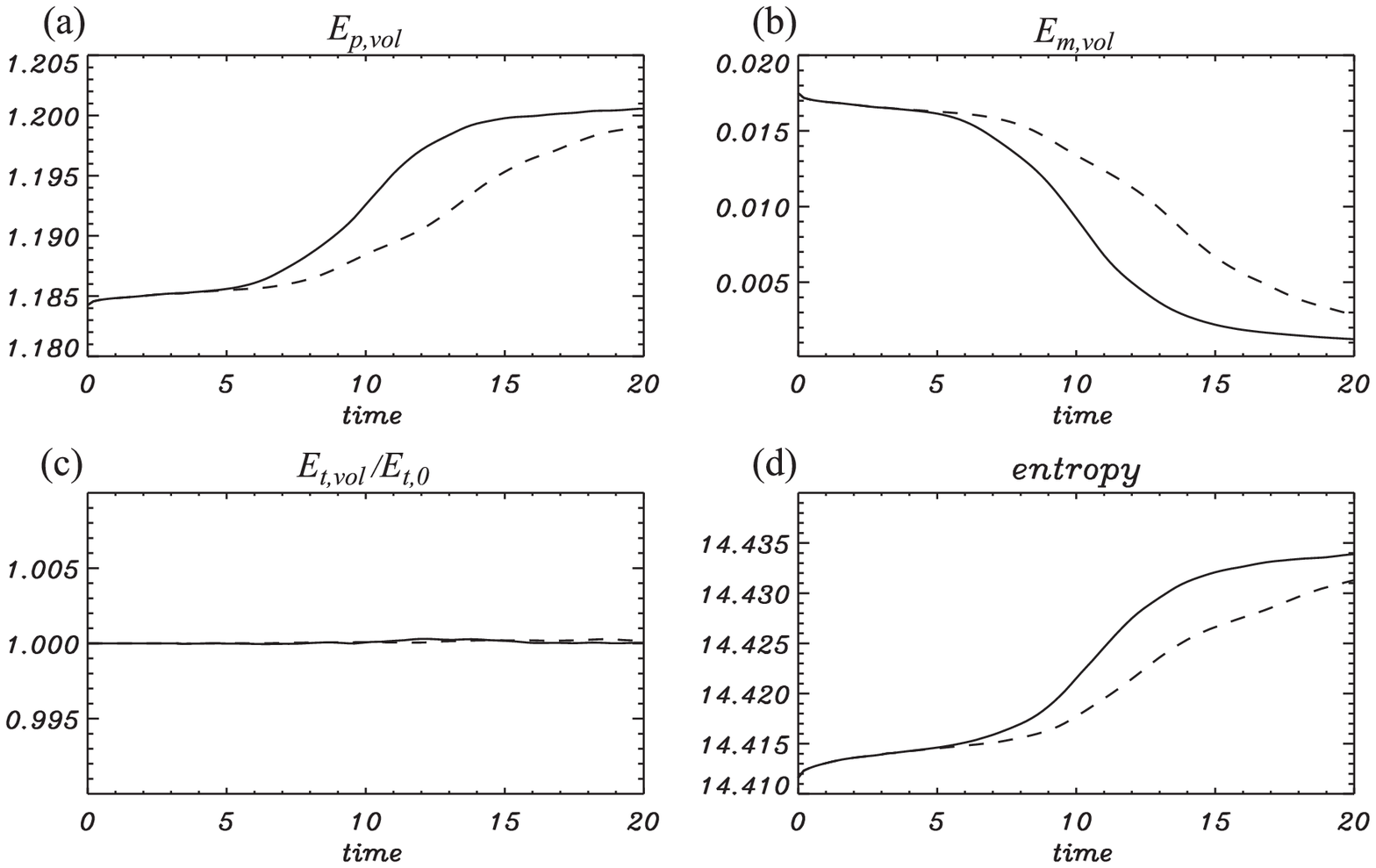}
\caption{Time evolution of the (a) volume-averaged plasma energy $E_{p}$, (b) magnetic energy $E_{m}$, (c) total energy $E_{t}$ normalized by initial total energy ($E_{t,0}$), and (d) entropy for case A ({\it solid lines}) and case B ({\it dashed lines}). 
\label{f2}}
\end{figure}
While the magnetic energy declines, the plasma energy increases gradually in the linear and rapidly in the nonlinear phase. Here growth of the CD kink instability leads to radial velocity increase which contributes a kinetic energy component to the plasma energy in both linear and non-linear phases. At about $t=11R/c$ for case A ($t=14R/c$ for case B), increase in the plasma energy nearly ceases and the hot plasma column is almost relaxed. The multiple mode case B shows a more gradual evolution in the nonlinear phase and later relaxation than the two mode case A. Multiple modes lead to a more gradual interaction, slower development of turbulent structure, and later relaxation of the hot plasma column. The volume-averaged total energy $E_{t}=E_{p}+E_{m}$ shown in Fig.\ 2c is conserved at the $0.01$\% level in both simulations. Figure 2d shows the time evolution of the volume-averaged entropy $S=\ln (p/\rho^{\gamma}) $. The entropy gradually increases in the linear phase and rapidly increases in the nonlinear phase.

Our results show that the magnetic field dissipates in the nonlinear phase. This occurs because strong gradients in the magnetic field are formed. In our simulations, dissipation is numerical; in the real world, one expects the same effect because the kink instability of a purely toroidal magnetic field develops at very small scales and as a result dissipation is inevitable. The volume averaged total energy, $E_{t}=E_{p}+E_{m}$, of the system (Fig.\ 2c) shows that variation of the total energy is at the 0.01\% level and is much less than the variation in the plasma energy. This indicates that our numerical dissipation conserves energy with sufficient accuracy.

\begin{figure}[h!]
\epsscale{0.5}
\plotone{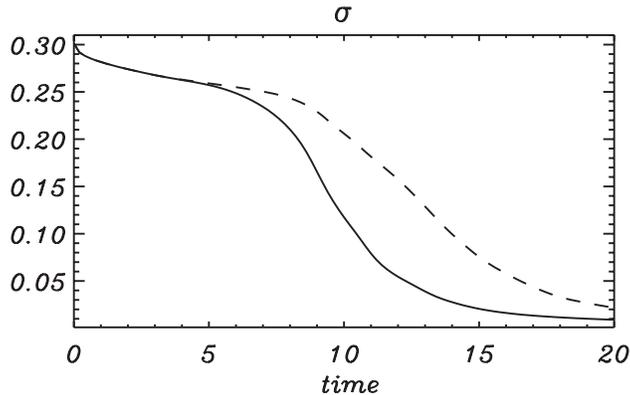}
\caption{Time evolution of the volume-averaged magnetization parameter $\sigma$ in the hot plasma column ($R \le 1$) for case A ({\it solid line}) and case B ({\it dashed line}). 
\label{f3}}
\end{figure}
Figure 3 shows time evolution of the magnetization parameter $\sigma=B^{2}/(\rho h)$ which averaged only over the magnetized hot plasma column but not over the whole volume for case A and B. Initially the volume-averaged magnetization $\sigma=0.3$ in the hot plasma column. In linear growth phase, $\sigma$ gradually decreases in both of cases. After $t=6R/c$ for case A ($t=7R/c$ for case B) $\sigma$ rapidly decreases because the magnetic field dissipates.

In order to study the relaxation of the hot plasma column in more detail, we show the time evolution of radial profiles of toroidal ($\phi$ direction) and axial ($z$ direction) averaged quantities in Figures 4 and 5. 
Initially the hot plasma column contains toroidal magnetic field in hydrostatic equilibrium. In the linear phase of the CD kink instability, the radial and axial components of the magnetic field grow while the toroidal magnetic field and gas pressure decline gradually beginning near the axis. When the CD kink instability enters the nonlinear phase (at $t=6R/c$ for case A and $t=7R/c$ for case B), the toroidal magnetic field and gas pressure decrease rapidly, and the radial and axial components of the magnetic field increase throughout the hot plasma column. At the end of the nonlinear phase (at $t=11R/c$ for case A and $t=14R/c$ for case B), all magnetic components become comparable and the field becomes totally chaotic. In a saturation phase, the magnetized column begins slow expansion so that eventually the column should merge with the surrounding gas. For the different initial perturbation profiles, the evolutionary time scale is different but the physical behavior is similar. Therefore, final relaxation of the hot plasma column is independent of the initial perturbation profile.
\begin{figure}[hp!]
\epsscale{0.8}
\plotone{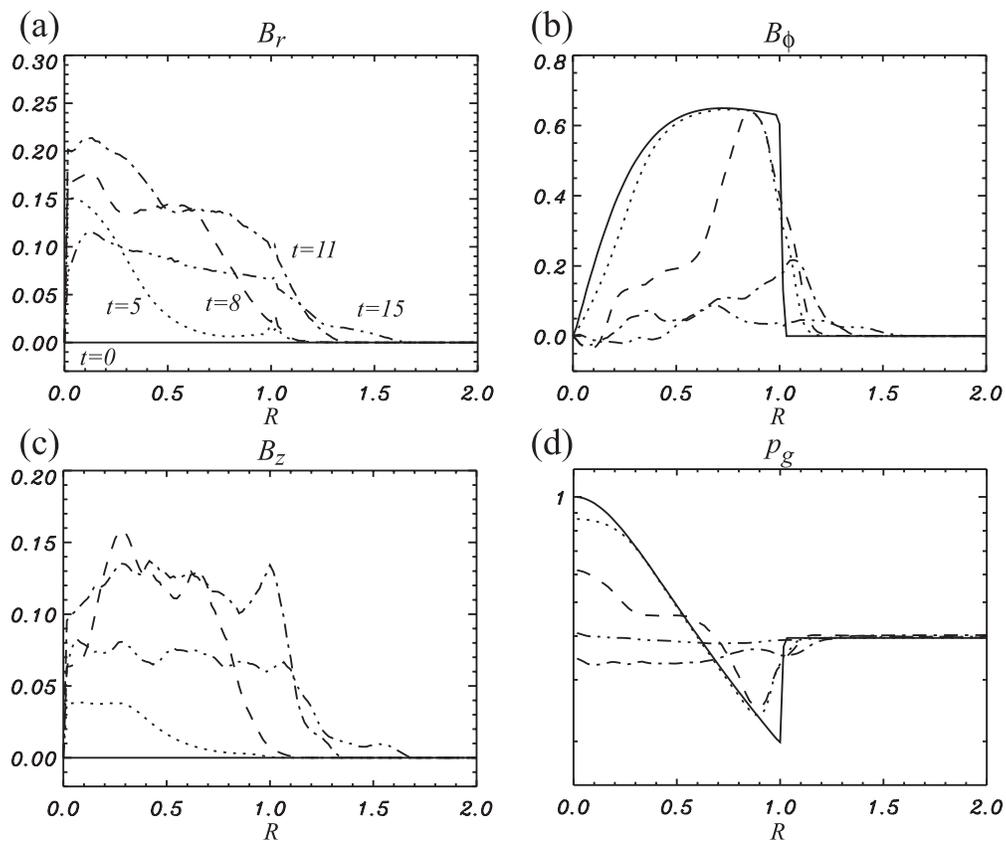}
\caption{Time evolution of the radial profile of the toroidal and axial-averaged radial ($B_{r}$), toroidal ($B_{\phi}$), and axial ($B_{z}$) magnetic field components and the gas pressure at $t=0R/c$ ({\it solid}), $5R/c$ ({\it dotted}), $8R/c$ ({\it dashed}), $11R/c$ ({\it dash-dotted}), and $15R/c$ ({\it dash-double-dotted}) for case A. 
\label{f4}}
\end{figure}
\begin{figure}[hp!]
\epsscale{0.8}
\plotone{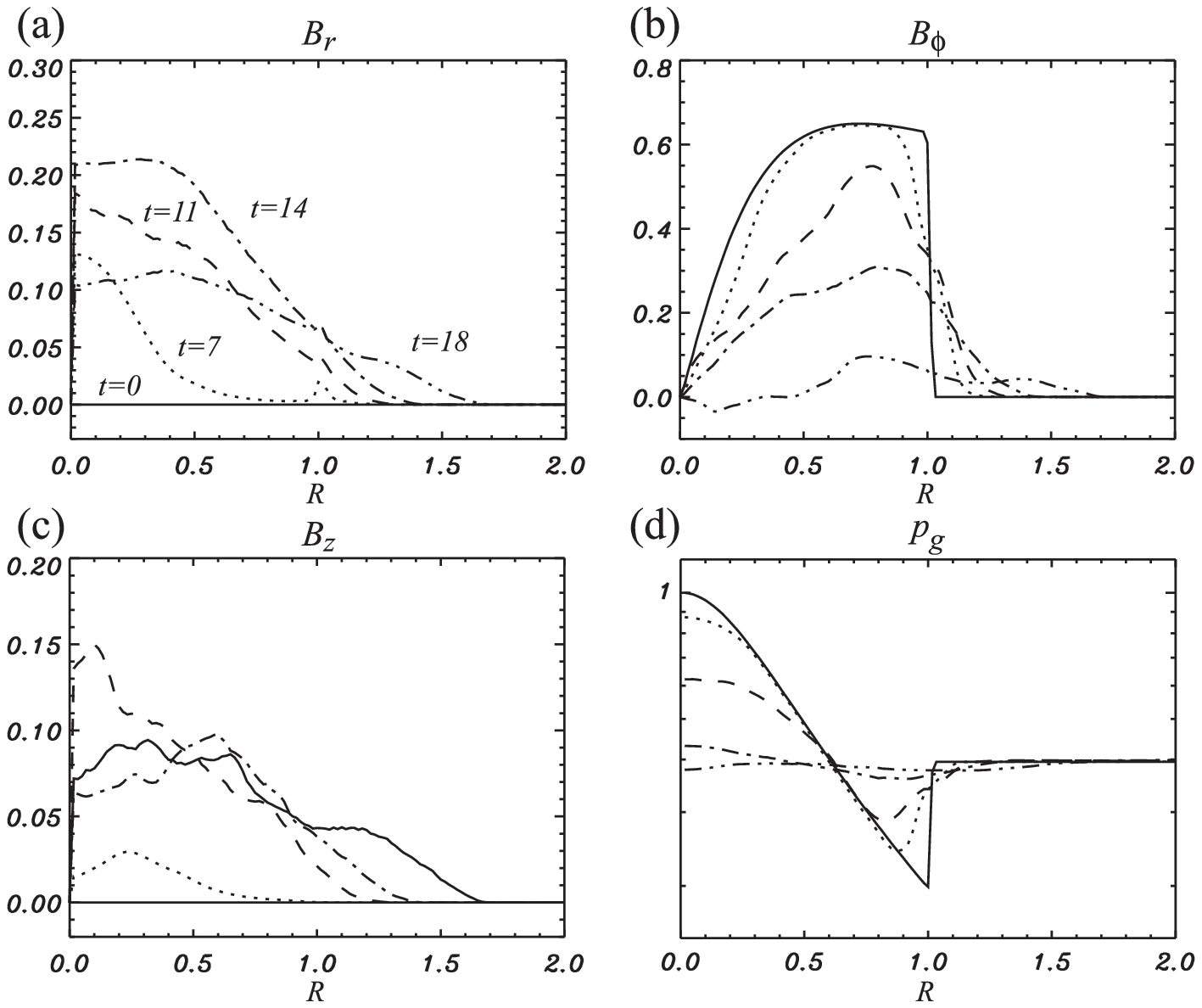}
\caption{Time evolution of the radial profile of the toroidal and axial-averaged radial ($B_{r}$), toroidal ($B_{\phi}$), and axial ($B_{z}$) magnetic field components and the gas pressure at $t=0R/c$ ({\it solid}), $7R/c$ ({\it dotted}), $11R/c$ ({\it dashed}), $14R/c$ ({\it dash-dotted}), and $18R/c$ ({\it dash-double-dotted}) for case B. 
\label{f5}}
\end{figure}

\section{Summary and Discussion}

We have investigated the development of the CD kink instability of a hydrostatic hot plasma column containing a toroidal magnetic field as a model for pulsar wind nebulae. The CD kink instability is excited by a small initial velocity perturbation and turbulent structure develops inside the hot plasma column. At the end of nonlinear evolution of the CD kink instability, the hot plasma column relaxes with a slow radial expansion. The magnetization $\sigma$ decreases from an initial value of $0.3$ to a final value of $0.01$. We find that for different initial perturbation profiles the time scale is a bit different but the physical behavior is the same. Therefore the relaxation of a hot plasma column is independent of the initial perturbation profile.

Our simulations confirm the scenario envisaged by Begelman (1998). Toroidal magnetic loops come apart, the hoop stress declines and the pressure difference across the nebula is washed out. In our simulations, the ratio of the gas pressure on the axis to the total (magnetic+gas) pressure at the plasma column boundary decreases from 2.5 to 1.5 during the linear phase, while magnetic dissipation is still small. In the nonlinear phase, the magnetic field dissipates and the gas pressure excess near the axis disappears. For this reason, elongation of a pulsar wind nebula cannot be correctly estimated by axisymmetrical models, because axisymmetric models  retain a concentric toroidal magnetic field geometry. 

Radiation from the Crab nebula is highly polarized along the axis of the nebula, which is indicative of a toroidal magnetic field. We see that even though the instability eventually destroys the toroidal structure, the magnetic field becomes completely chaotic only at the end of the nonlinear stage of development. Therefore, the toroidal magnetic field should dominate in the central parts of the nebula that are filled by recently injected plasma. 

Our simple model does not allow us to determine the value of $\sigma$ in the pulsar wind; for this purpose one has to perform fully 3D simulations of the expanded nebula taking into account the continuous injection of  plasma. Here we have demonstrated that 3D effects are crucially important to a determination of the structure of pulsar wind nebulae and that previous dynamical arguments concluding that $\sigma$ must be extraordinarily small can be abandoned.

\acknowledgments
This work has been supported by NSF awards AST-0506719, AST-0506666, AST-0908010, and AST-0908040, and NASA awards NNG05GK73G, NNX07AJ88G, and NNX08AG83G, and US-Israeli BSF award 2006170. The simulations were performed on the Columbia Supercomputer at the NAS Division of the NASA Ames Research Center, the SGI Altix (cobalt) at the National Center for Supercomputing Applications in the TeraGrid project supported by the NSF, and the Altix3700 BX2 at YITP in Kyoto University.


\begin{thebibliography}{}

\bibitem[Arons(2007)]{Aro07} Arons, J. 2008, in Neutron Stars and Pulsars, 40 Years After the Discovery, ed. W. Becker \& H. H. Huang (MPE Rep. 291; Garching; MPE) (arXiv:0708.1050) 

\bibitem[Begelman(1998)]{Beg98} Begelman, M. C. 1998, \apj, 493, 291

\bibitem[Begelman \& Li(1992)]{Bel92} Begelman, M. C., \& Li, Z.-Y. 1992, \apj, 397, 187

\bibitem[Bogovalov(1999)]{Bog99} Bogovalov, S. V. 1999, \aap, 349, 1017

\bibitem[Camus et al.(2009)]{Cam09} Camus, N. F., Komissarov, S. S., Bucciantini, N., \& Hughes, P. A. 2009, \mnras, 400, 1241

\bibitem[Del Zanna et al.(2004)]{Del04} Del Zanna, L., Amato, E., \& Bucciantini, N. 2004, \aap, 421, 397

\bibitem[Del Zanna et al.(2006)]{Del06} Del Zanna, L., Volpi, D., Amato, E., \& Bucciantini, N. 2006, \aap, 453, 621

\bibitem[Emmering \& Chevalier(1987)]{Emm87} Emmering, R. T., \& Chevalier, R. A. 1987, \apj, 321, 334

\bibitem[Gaensler \& Slane(2006)]{Gae06} Gaensler, B., \& Slane, P. 2006, \araa, 44, 17

\bibitem[Jiang \& Shu(1996)]{Jia96} Jiang, G. S., \& Shu, C. W. 1996, J. Comput. Phys., 126, 202

\bibitem[Kennel \& Coroniti(1984a)]{Ken84a} Kennel, C. F., \& Coroniti, F. V. 1984a, \apj, 283, 694

\bibitem[Kennel \& Coroniti(1984b)]{Ken84b} Kennel, C. F., \& Coroniti, F. V. 1984b, \apj, 283, 701

\bibitem[Kirk \& Skjaeraasen(2003)]{Kir03} Kirk, J. G., \& Skjaeraasen, O. 2003, \apj, 591, 366

\bibitem[Kirk et al.(2009)]{Kir09} Kirk, J. G., Lyubarsky, Y., \& Petri, J. 2009, in Neutron Stars and Pulsars, ed. W. Becker, Astrophysics and Space Science Library, 357, 421.

\bibitem[Komissarov(1999)]{Kom99} Komissarov, S.S. 1999, \mnras, 303, 343

\bibitem[Komissarov \& Lyubarsky(2003)]{Kom03} Komissarov, S., \& Lyubarsky, Y. 2003, \mnras, 344, L93

\bibitem[Komissarov \& Lyubarsky(2004)]{Kom04} Komissarov, S., \& Lyubarsky, Y. 2004, \mnras, 349, 779

\bibitem[Lyubarsky(2003)]{Lyu03} Lyubarsky, Y. E. 2003, \mnras, 345, 153

\bibitem[Lyubarsky \& Kirk(2001)]{Lyu01} Lyubarsky, Y., \& Kirk, J. G. 2001, \apj, 547, 437

\bibitem[Michel(1971)]{Mic71} Michel, F. C. 1971, Comments Astrophys. Space Phys., 3, 80
\bibitem[Mizuno et al.(2006)]{Miz06} Mizuno, Y., Nishikawa, K.-I.,
Koide, S., Hardee, P., \& Fishman, G. J. 2006, ArXiv Astrophysics e-prints, 0609004

\bibitem[Mizuno et al.(2009)]{Miz09} Mizuno, Y., Lyubarsky, Y., Nishikawa, K.-I., \& Hardee, P. E. 2009, \apj, 700, 684

\bibitem[Mizuno et al.(2010)]{Miz10} Mizuno, Y., Hardee, P. E., \& Nishikawa, K.-I. 2010, \apj, submitted

\bibitem[Petri \& Lyubarsky(2007)]{Per07} P\'{e}tri, J., \& Lyubarsky, Y. 2007, \aap, 473, 683

\bibitem[Rees \& Gunn(1974)]{Ree74} Rees, M. J., \& Gunn, J. E. 1974, \mnras, 167, 1

\bibitem[Spitkovsky(2006)]{Spi06} Spitkovsky, A. 2006, \apj, 648, L51

\bibitem[Volpi et al.(2008)]{Vol08} Volpi, D., Del Zanna, L., Amato, E., \& Bucciantini, N. 2008, \aap, 485, 337

\bibitem[Zenitani \& Hoshino(2007)]{Zen07} Zenitani, S., \& Hoshino, M. 2007, \apj, 670, 702

\end{thebibliography}
\end{document}